\documentclass[%
preprintnumbers,
nofootinbib,
 amsmath,amssymb,
 aps,
 showkeys,
]{revtex4-2}

\bibliography{report}

\usepackage{graphicx,color}
\usepackage{dcolumn}
\usepackage{bm}

\begin{document}

\title{Extraordinary nature of the nucleon scalar charge and its densities\\
as a signal of nontrivial vacuum structure of QCD}

\author{Masashi Wakamatsu}
 \email{wakamatu@post.kek.jp}
\affiliation{%
KEK Theory Center, Institute of Particle and Nuclear Studies,
High Energy Accelerator Research Organization (KEK),
Oho 1-1, Tsukuba, 305-0801, Ibaraki, Japan
}%




\date{\today}

\begin{abstract}
It is widely known that the nucleon scalar charge is proportional 
to the pion-nucleon sigma term as one of the important low energy
observables of QCD. Especially interesting to us is the physics of the nucleon 
scalar charge densities. This comes from the fact that the corresponding
operator has the same quantum number as the physical vacuum.
It indicates unusual behavior of the the nucleon 
scalar density as a function of the distance $r$ from the nucleon center.
Namely, it would not be reduced down to zero at the spatial infinity but rather
approaches some nonzero constant corresponding to the vacuum quark condensate.
Naturally, this unique nature of the nucleon scalar density in the
position space also affects the corresponding density in the momentum space,
i.e. the corresponding parton distribution function (PDF) as a function 
of the Bjorken variable $x$. This PDF is known as the chiral-odd twist-3 PDF
$e (x)$. We argue that $e(x)$ is likely to have a delta-function type 
singularity at $x=0$, and that the appearance of this singularity
can be interpreted as a signal of the nontrivial vacuum structure of the QCD.
\end{abstract}

\keywords{Time-dependent Aharonov-Bohm effect, 
4-dimensional Stokes theorem, quantum mechanics, gauge transformation}


\pacs{01.55.+b, 03.65.-w, 03.65.Vf,11.15.-q}

\maketitle


\section{Introduction}

\vspace{1mm}
It is well-known that the nucleon scalar charge is related to the pion-nucleon sigma term 
which is one of the important low energy observables \cite{Jameson1992, Sainio2001}.
However, since the standard model of elementary particles is a $V$-$A$ (vector and axial-vector) theory, 
there is no external electro-weak current which directly couples to the nucleon scalar charge as well as 
to the tensor charge \cite{Cottingham-Greenwood2007, Schwartz2014}.
In recent years, these quantities attract wide interest in search of physics beyond the standard model, 
which also allows $S$-$T$ (scalar and tensor) couplings \cite{Courtoy2015, Courtoy2018}.
In the present paper, we demonstrate that far more interesting than the nucleon scalar charge
itself is its densities in the coordinate space as well as in the momentum space.
Why are they interesting ? The ultimate reason can be traced back to the fact that the 
corresponding operator has the same quantum number as the {\it physical vacuum} of QCD.
As is widely known, as a consequence of spontaneous chiral-symmetry breaking ($\chi SB$), 
the QCD vacuum is believed to be 
characterized by non-zero quark condensate, i.e. non-zero scalar quark density.
This implies that the nucleon scalar density as a function of the distance $r$ from the nucleon center is 
expected to show the following abnormal behavior.
Namely, as the distance $r$ increases, the nucleon scalar density does not attenuate to zero, 
but it would rather approach nonzero value corresponding to the vacuum quark condensate.
A natural question is whether this unique nature of the nucleon scalar charge density would show up 
somewhere in physical observables.
The purpose of the present concise review is to show that we can answer the above question
affirmatively.

\section{Physics behind the nucleon scalar charge}

There already exist several lattice QCD calculations of the pion-nucleon sigma term $\Sigma_{\pi N}$ 
or the nucleon scalar charge $\bar{\sigma}$ \cite{Yamanaka2018, Hasan2019, Alexandrou2020}. 
They are related as $\Sigma_{\pi N} = m_{ud} \,\bar{\sigma}$,
where $m_{ud}$ is the average mass of the up and down quarks.
To get a rough idea about the magnitude of $\bar{\sigma}$ or $\Sigma_{\pi N}$, here we
quote the results of the recent lattice QCD simulation by Alexandrou et al. \cite{Alexandrou2020}.
(Note that the scalar coupling $g_S$ in their notation is identified with the nucleon scalar 
charge $\bar{\sigma}$ in our notation.) 
Their prediction for $\bar{\sigma}$ is given as a sum of four terms : 
\begin{equation}
 \bar{\sigma} \ = \ \bar{\sigma} (u + d (conn)) \ + \ \bar{\sigma} (u + d (disc)) 
 \ + \ \bar{\sigma} (s)  \ + \ \bar{\sigma} (c) ,
\end{equation}
with
\begin{eqnarray}
 &\,& \bar{\sigma} (u + d \,(conn)) \ = \ 20.4 (1.6), \ \ \ \bar{\sigma} (u + d \,(disc)) \ = \ 3.04 (59) , \\ 
 &\,& \bar{\sigma} (s) \ = \ 1.00 (13), \ \ \ \bar{\sigma} (c) \ = \ 0.175 (36) .
\end{eqnarray}
Here, $\bar{\sigma} (u + d \,(conn))$ and  $\bar{\sigma} (u + d \,(disc))$ respectively stands for
the contribution of the connected and disconnected diagrams coming from the up and
down quarks, while $\bar{\sigma} (s)$ and $\bar{\sigma} (c)$ represent the
contributions from the strange and charm quarks.
One sees that the contribution of the connected diagrams dominates over that of
the disconnected diagrams. However, it should be kept in mind that the
separation into the connected and disconnected pieces in the lattice QCD simulation
does not necessarily correspond to directly observable separation.    
The final prediction for the pion-nucleon sigma term is also given as \cite{Alexandrou2020}

\begin{equation}
 \Sigma_{\pi N} \ = \ m_{ud} \,\bar{\sigma} \ \simeq \ 41.6 \,(3.8) \,\mbox{MeV} ,
\end{equation}
although the value of $m_{ud}$ is not explicitly written in their paper.
Anyhow, this value seems roughly consistent with the empirical one obtained from
the analysis of the low energy pion-nucleon scattering data 
\cite{GLS1991, HEKM2015, HEKM2023}.

To understand the fundamental importance of the pion-nucleon sigma term, 
it would be useful to briefly  recall the physics behind the nucleon scalar charge.
First, let us remember the theoretical prediction of the MIT bag model as
a prototype low energy effective theory of the nucleon \cite{Johnson1975}.
Its prediction is given as follows, 
\begin{equation}
 \bar{\sigma} \ = \ \langle N \,\vert \,\int \,\bar{\psi} (\bm{r}) \,\psi (\bm{r}) \,d^3 r \,
 \vert N \rangle \ = \  N_c \,\int _0^\infty \,\left\{ f(r)^2 \ - \ g (r)^2 \right\} \,r^2 \,d r.
\end{equation}
Here, $N_c = 3$ is the number of colors, while $f (r)$ and $g(r)$ are the upper and
lower components of the radial wave function in the MIT bag model.
Since the radial wave functions are normalized as
\begin{equation}
 \int_0^\infty \,\left\{ f(r)^2 \ + \ g(r)^2 \right\} \,r^2 \,d r \ = \ 1 , 
\end{equation}
and since the upper component $f(r)$ dominates over the lower component $g(r)$,
we are inevitably led to an inequality,
\begin{equation}
 \int_0^\infty \,\left\{ f(r)^2 \ -\ g (r)^2 \right\} \,r^2 \,d r \ < \ 1.
\end{equation}
This in turn leads to the remarkable inequality
\begin{equation}
 \bar{\sigma} \ < \ N_c \ = \ 3.
\end{equation}
For reasonable choice of the quark mass $m_{ud}$, this gives too small 
$\Sigma_{\pi N}$, which is largely incompatible
with the existing empirical information for the pion-nucleon sigma term \cite{Bernard1996}.
We must therefore conclude that naive quark model with only three valence quark 
degrees of freedom sizably underestimates the magnitude of $\Sigma_{\pi N}$.

\begin{figure}[h]
\begin{centering}
\includegraphics[width=8.0 cm]{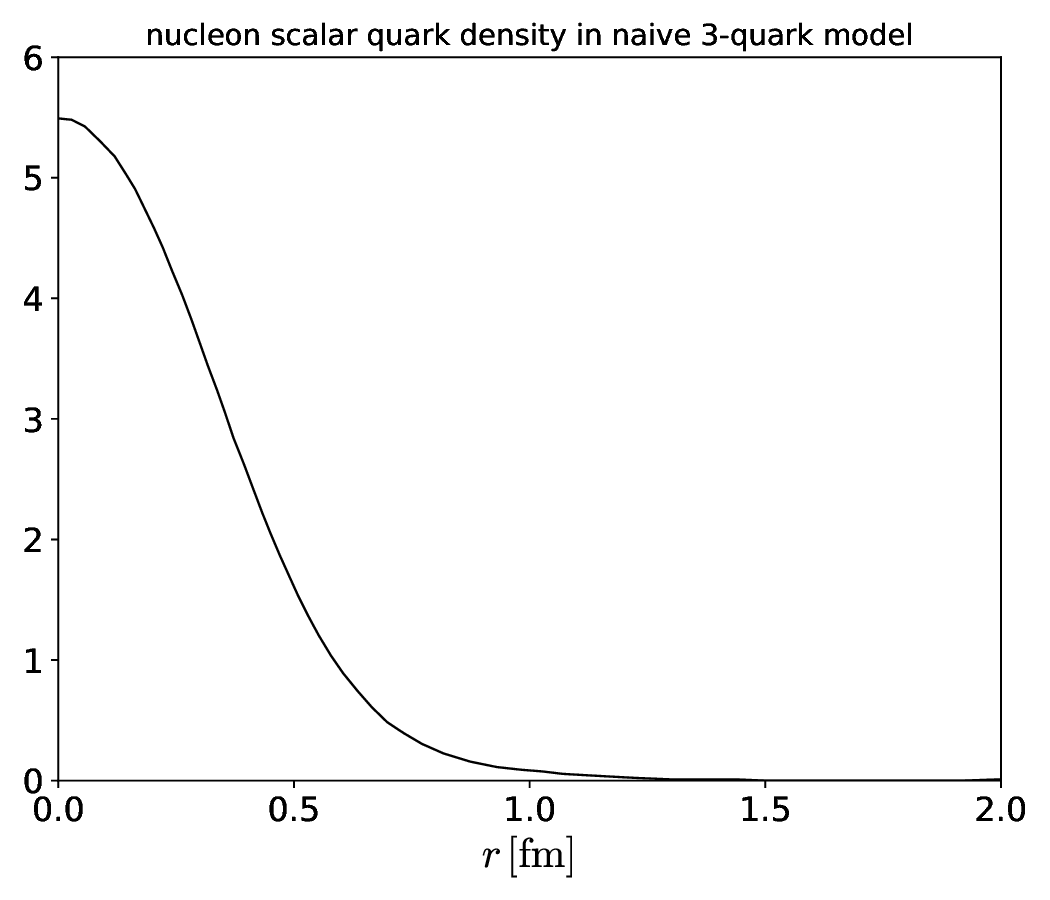}
\caption{Typical prediction of the naive three-quark model for the nucleon scalar charge
density $\rho_S (r)$ in the coordinate space.\label{fig1}}
\end{centering}
\end{figure}

\vspace{3mm}
A fatal shortcoming of naive three-quark models in the scalar channel is also clear 
from its prediction for the nucleon scalar charge density in the coordinate space.
A typical prediction of the naive three quark model of the nucleon for the
scalar density is illustrated in Fig.\ref{fig1}. 
(The scalar quark density $\rho_S (r)$ shown in Fig.\ref{fig1} is normalized as
$4 \pi \int_0^\infty \,\rho_S (r) \,r^2 \,d r = \bar{\sigma}$, with $\bar{\sigma}$ being
the nucleon scalar charge. Since the unit of $r$ is given by
fm (fermi or femtometre) , the unit of $\rho_S (r)$ here is $\mbox{fm}^{-3}$.)
As one sees, the scalar density takes a maximum
value at the center of the nucleon and it smoothly attenuates to zero as the distance $r$
from the nucleon center becomes large and approaches infinity.
One should recognize that this contradicts our expectation that, at least in the region
far apart from the the nucleon center, the scalar quark density must coincide with
the nonzero value of the vacuum quark condensate, as long as we believe the scenario 
that the spontaneous breaking on the chiral symmetry generates nonzero vacuum
quark condensate \cite{Brodsky2010}. 
As we shall see in the next section, the chiral quark soliton model (CQSM) is 
a very unique effective model of the nucleon, which can simultaneously reproduce 
the nontrivial vacuum 
quark condensate and the local structure of the nucleon scalar charge density. 

\unskip

\vspace{3mm}
\section{Brief introduction to the chiral quark soliton model}

\vspace{1mm}
The chiral quark soliton model (CQSM) is a low energy effective model of
baryons first introduced by Diakonov et al. based on the instanton-liquid
picture of the QCD vacuum \cite{DPP1988}.
The effective Lagrangian of the CQSM is given by
\begin{equation}
 {\cal L}_{CQSM} \ = \ \bar{\psi} (x) \,\left( i \,\gamma^\mu \,\partial_\mu \ - \ M \,
 e^{\ i \,\gamma_5 \,\bm{\tau} \cdot \bm{\pi} (x) / f_\pi} \right) \,\psi (x),
\end{equation}
where $\psi (x)$ and $\bm{\pi} (x)$ represent the effective quark and pion fields,
while $M$ stands for the effective quark mass (or the constituent quark mass)
of the order of $400 \,\mbox{MeV}$. 
Note that there is no kinetic term for the pion in this lagrangian, which 
implies that the pion field in this model is not an independent field but it a 
dependent field of quarks (or the quark-antiquark composite). 
The effective pion action $S_{eff} [\bm{\pi}]$
obtained from this lagrangian can be defined by formally carrying out
the path integration over the quark field,
\begin{equation}
 Z \ = \ \int {\cal D} \bm{\pi} \int {\cal D} \psi \,{\cal D} \psi^\dagger \,\,
 e^{\,\,i \,\int d^4 x \,{\cal L}_{CQSM}} \ = \ 
 \int {\cal D} \bm{\pi} \,\,e^{\,\,i \,S_{eff} [\bm{\pi}]} .
\end{equation}
It is known that, if we use the derivative-expansion type approximation
in the three flavor case, we obtain an effective meson action of the following structure, 
\begin{eqnarray}
 S_{eff} [\bm{\pi}] &=& \mbox{Skyrmion action with Wess-Zumino term} \nonumber \\
 &+& \mbox{destabilizing 4-th derivative term} \nonumber \\
 &+& \cdots .
\end{eqnarray}
Unfortunately, different from the original Skyrmion action, the existence of the
destabilizing 4-th derivative term does not allow existence of stable soliton-like solution.

\begin{figure}[h]
\begin{centering}
\includegraphics[width=4.6cm]{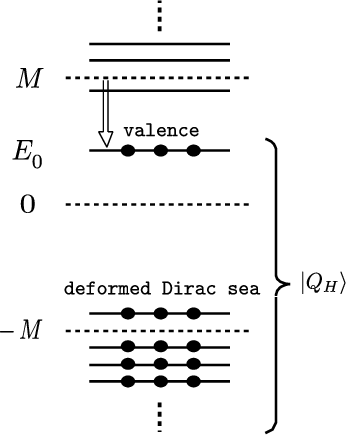}
\caption{Characteristic behavior of the single-quark energy levels under
the mean-field of hedgehog shape.\label{fig2}}
\end{centering}
\end{figure}

\vspace{2mm}
The basic idea of the CQSM is to construct a stable soliton-like localized solution 
{\it without} relying upon derivative-expansion type approximation.
Basically, it is a relativistic mean-field theory for quark fields.
We start with the assumption that the pion field, which plays the role of mean field
for quarks, takes the hedgehog form as follows similar to the famous 
Skyrme model,
\begin{equation}
 \bm{\pi} (\bm{r}) \ = \ \hat{\bm{r}} \, F (r) ,
\end{equation}
where the function $F (r)$ is supposed to satisfy the following boundary condition,
\begin{equation}
 F (0) \ - \ F (\infty) \ = \ n \,\pi
\end{equation}
with $n \,( = 1)$ being so-called the winding number of the effective pion field.
Under the presence of this mean field, the quark field obeys the following 
Dirac equation,
\begin{equation}
 H \,\vert m \rangle \ = \ E_m \,\vert m \rangle,
\end{equation}
with
\begin{equation}
 H \ = \ \frac{\bm{\alpha} \cdot \nabla}{i} \ + \ M \,\beta \,
 \left( \cos F (r) \ + \ i \,\gamma_5 \,\sin F (r) \right) .
\end{equation}
A characteristic feature of this Dirac equation with the topologically twisted
hedgehog mean field is that one deep single-quark bound state
appears from the positive energy continuum of the above Dirac Hamiltonian. 
 (See Fig.\ref{fig2} for illustration.)
We call this particular single-quark level the {\it valence quark orbital}. 

\unskip

\vspace{3mm}
An object having baryon number one with respect to the
physical vacuum is obtained by putting $N_c \,(= 3)$ quarks into this
valence orbital as well as all the negative energy (Dirac-sea)
orbitals. This baryon number one object with respect to the physical vacuum 
is sometimes called the quark hedgehog denoted as $\vert Q_H \rangle$.
Accordingly, the total energy of this quark hedgehog is given as
a sum of the valence quark contribution and the vacuum polarization
contribution as
\begin{equation}
 E_{static} \ = \ N_c \,E_0 \ + \ E_{v.p.} ,
\end{equation}
where $E_0$ is the single-particle energy of the valence quark level,
while the vacuum polarization contribution represents the
Casimir energy resulting from the polarization (deformation)
of the Dirac-sea quark orbitals and given by
\begin{equation}
 E_{v.p.} \ = \ N_c \,\left( \sum_{m \,(E_m < 0)} \,E_m \ - \ 
 \sum_{k \,(\epsilon_k < 0)} \epsilon_k \right). \label{Eq:E_vp}
\end{equation}
That is, the Casimir energy is given as a sum of all the energies
of quarks in the negative-energy Dirac sea orbitals. 
Here in Eq.(\ref{Eq:E_vp}) , the 2nd term represents the subtraction of 
the Dirac sea energy of the physical vacuum. (The physical vacuum of the model is
obtained by letting $F (r) \rightarrow 0$.)
The most probable pion field configuration is then determined on
the basis of the stationary requirement for the total energy $E_{static} [F(r)]$,
\begin{equation}
 \frac{\delta}{\delta F(r)} \,E_{static} [ F (r)] \ = \ 0.
\end{equation}
This requirement combined with the above Dirac
equation is reduced to a self-consistent Hartree problem
which can be solved by the numerical method of Kahana and Ripka \cite{KR1984}.
(See \cite{WY1991} for more detail about the actual calculation method.) 
After self-consistency is fulfilled, the hedgehog pion field,
which was originally introduced as an external mean field for
quarks, becomes an implicit functional of the quark fields.

\vspace{2mm}
Actually, the vacuum polarization energy given by Eq.(\ref{Eq:E_vp}) contains
ultraviolet (logarithmic) divergence.
Often, this ultraviolet divergence is removed with the use of the Pauli-Villars 
regularization, which means the following replacement of  the effective action 
\begin{equation}
 S_{eff} [\bm{\pi}] \ \rightarrow \ S^M_{eff} [\bm{\pi}] \ - \ 
 \left( \frac{M}{M_{PV}} \right)^2 \,
 S^{M_{PV}}_{eff}  [\bm{\pi}] ,
\end{equation}
where $M_{PV}$ is a Pauli-Villars cutoff mass. However, since
the vacuum quark condensate contains {\it quadratic divergence}, the single 
subtraction is not enough and we need double-term Pauli-Villars subtraction
as used in \cite{KWW1999}, 
\begin{equation}
 S_{eff} [\bm{\pi}] \ \rightarrow \ S^M_{eff} [\bm{\pi}] \ - \ 
 \sum_{i = 1}^2 \,c_i \, S^{\Lambda_i}_{eff}  [\bm{\pi}] ,
\end{equation}
The four subtraction parameters $c_1, c_2, \Lambda_1, \Lambda_2$  are determined 
so as to remove quadratic and logarithmic divergence of the effective action and 
to reproduce the empirical value of vacuum condensate and correct pion kinetic
energy term in the effective pion action \cite{KWW1999}. 
Once these parameters are fixed, the model is known to reproduce low energy
observables of the nucleon as well as the various quark distributions of the
nucleon remarkably well  \cite{DPPPW1996, DPPPW1997, WGR1997L, WGR1997,
WK1998, Waka2003A, Waka2003B}.

\begin{figure}[h]
\begin{centering}
\includegraphics[width=7.5cm]{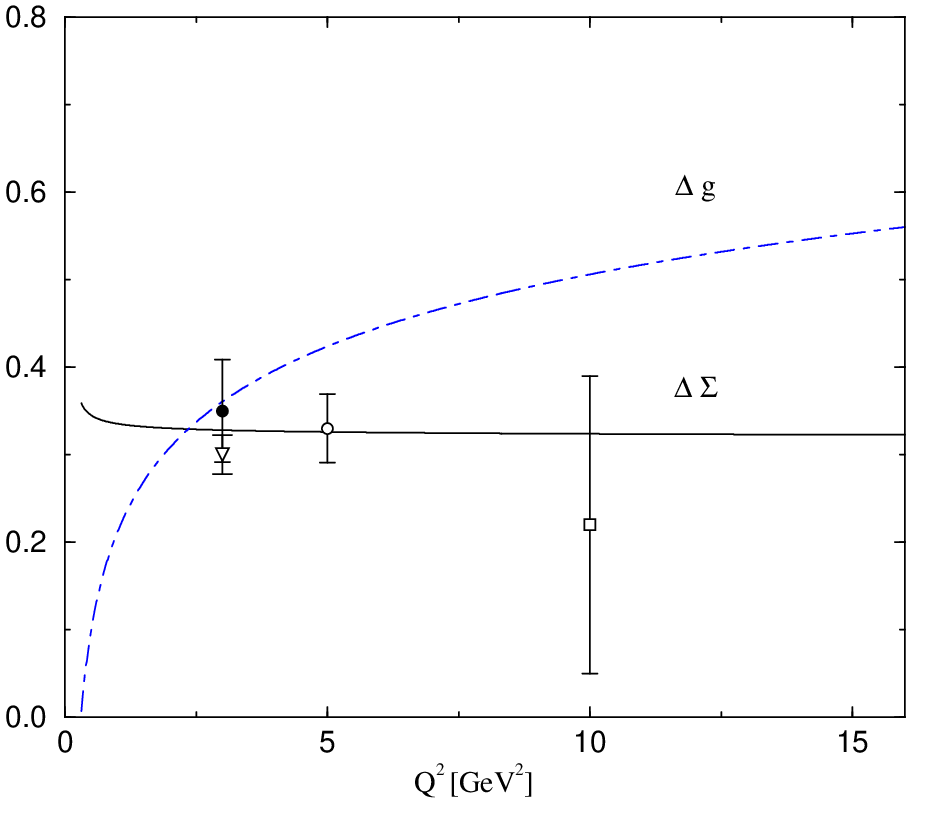}
\caption{The CQSM prediction of quark spin $\Delta \Sigma$ \cite{Waka2007}
as compared with the old experimental  data from the SMC group \cite{SMC1998}
and the newer data from the COMPASS \cite{COMPASS2005, COMPASS2007} and 
HERMES group \cite{HERMES2007}. \label{fig3}}
\end{centering}
\end{figure}

\vspace{2mm}
To convince the reliability of the CQSM, we show below its characteristic 
predictions related to the most important parton distribution functions 
at the twist-2 level. Probably, one of the remarkable
predictions of the CQSM is that it reproduces the observed small quark spin 
contribution  to the total nucleon spin fairly well. 
We show in Fig.\ref{fig3} the prediction of the CQSM for the longitudinal quark spin
$\Delta \Sigma$ and the longitudinal gluon spin $\Delta g$ in the nucleon 
as compared with the empirical information. 
Here, the scale dependencies of $\Delta \Sigma$ and $\Delta g$ are taken into 
account by using the evolution (DGLAP) equation at the next-leading order (NLO) 
under the assumption that $\Delta g = 0$ at the initial energy scale 
$Q^2_{ini} = 0.30 \,\mbox{GeV}^2$, which we identify the energy scale of 
our effective quark model.
One sees that the CQSM predicts fairly small quark spin contents in the nucleon
and it is qualitatively consistent with the empirical information. 
We emphasize that the small prediction of the CQSM for $\Delta \Sigma$ is 
deeply connected with its nucleon picture as a rotating hedgehog object. 
The time-dependent rotation of the hedgehog mean field necessarily enhances 
the contribution of the quark orbital angular momentum, which in turn reduces
the contribution of the intrinsic quark spin.  

\begin{figure}[h]
\begin{centering}
\includegraphics[width=12.0cm]{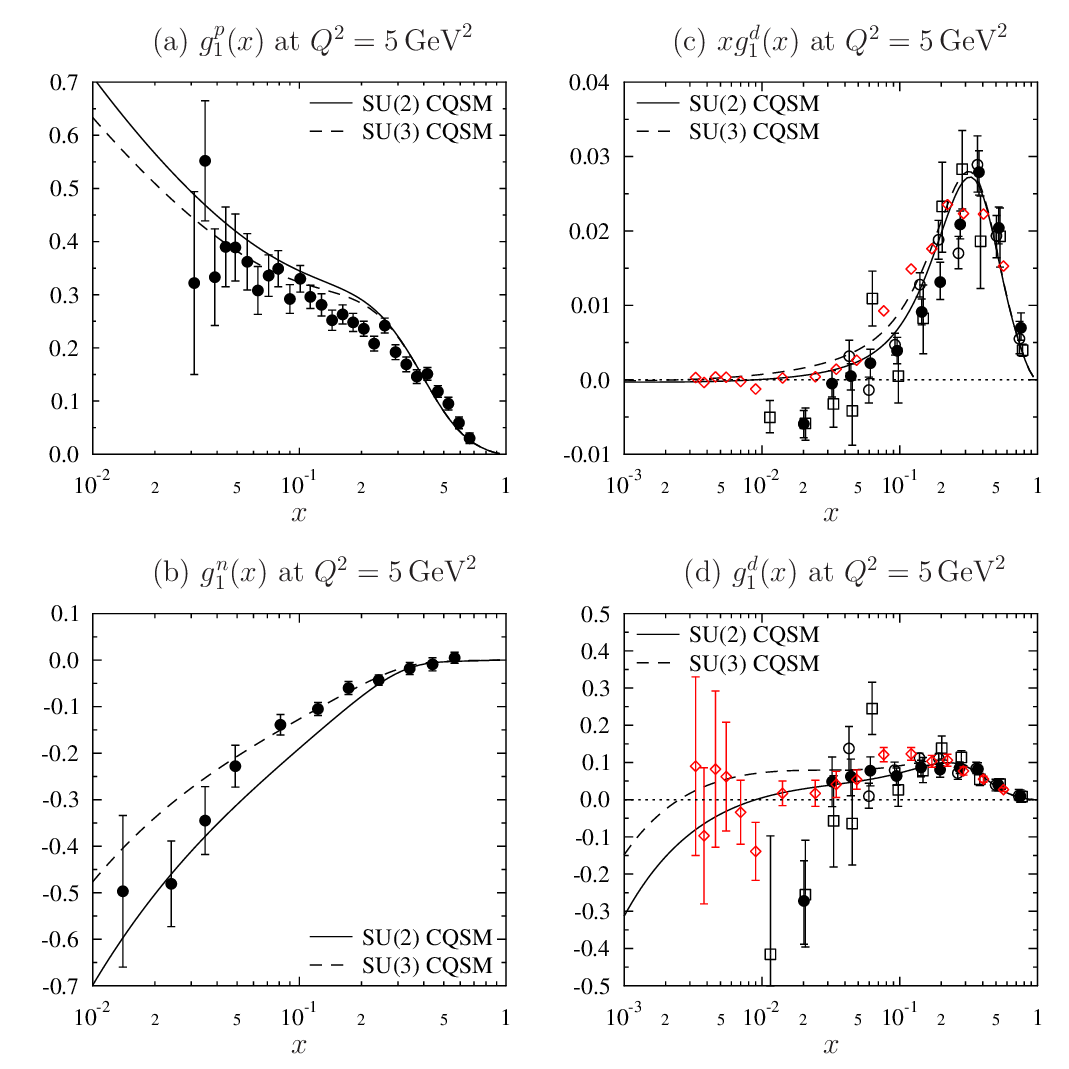}
\caption{The CQSM prediction for the longitudinally polarized structure functions
for the proton, the neutron and the deuteron as compared with
the old experimental  data from the SMC group \cite{SMC1998} and the newer data 
(red in color) from the COMPASS \cite{COMPASS2005, COMPASS2007}. 
For reference, the prediction of the flavor SU(3) version of the CQSM
is also shown. \label{fig4}}
\end{centering}
\end{figure}

\vspace{4mm}
Next, shown in Fig.\ref{fig4} are the predictions of the CQSM for the longitudinally
polarized structure functions of the proton, the neutron and the deuteron
as compared with the corresponding experimental data by the SMC group \cite{SMC1998}
and the Compass group \cite{COMPASS2005, COMPASS2007} and 
HERMES group \cite{HERMES2007}.  We can say that the agreement with the empirical 
information is encouraging especially in view of the fact that the predictions of the 
CQSM are almost parameter free.

\begin{figure}[h]
\begin{centering}
\includegraphics[width=12.0cm]{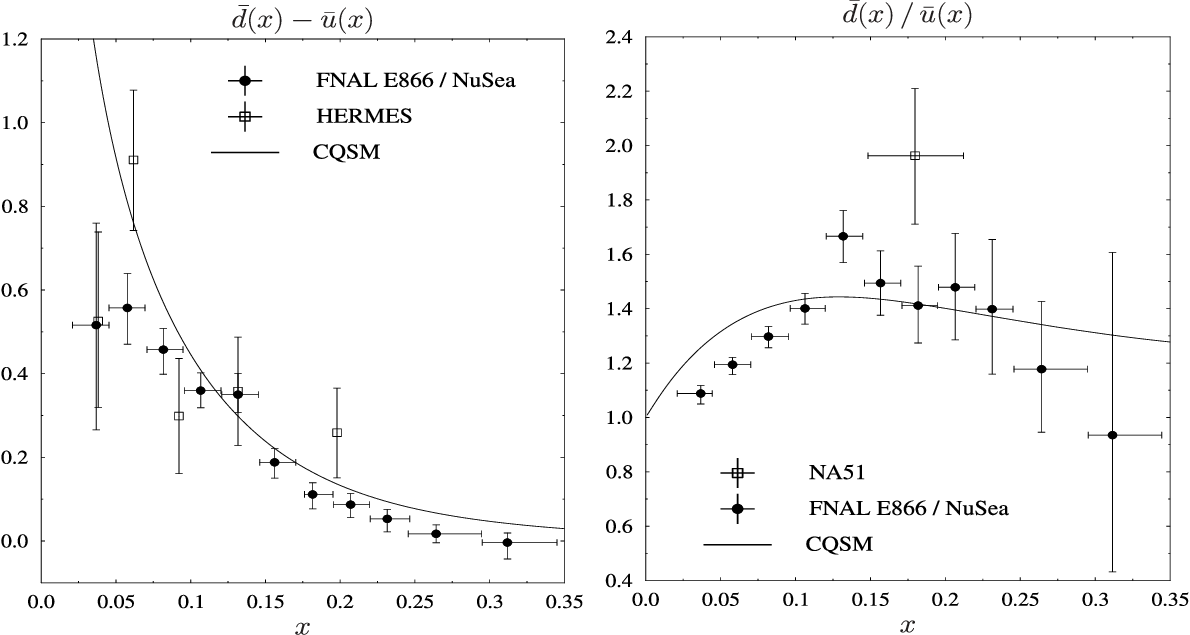}
\caption{the predictions
of the CQSM for the $\bar{d} (x) - \bar{u} (x)$ distribution and
the $\bar{d} (x) \,/\, \bar{u} (x)$ in comparison with the corresponding
experimental data from the Hermes \cite{HERMES1998}, 
FNAL-E866/NuSea group \cite{NuSea2001}
as well as the old data from NA51 \cite{NA51}. \label{fig5}}
\end{centering}
\end{figure}

\vspace{4mm}
Still another prominent feature of the CQSM is that it can give reliable predictions
about the sea-quark distributions or the anti-quark distributions  in the nucleon.
This greatly owes to its field theoretical nature, which takes account of the 
deformation (or the vacuum polarization) of the Dirac sea under the presence of
the hedgehog mean field in a nonperturbative manner.
It is empirically known that the distribution functions of the anti-quarks in
the proton is not flavor symmetric, i.e.  the distribution of the $\bar{d}$-quark   
dominates over that of the $\bar{u}$-quark inside the proton.
It is widely known that this flavor asymmetry of anti-quark distribution can be
explained by the effects of pion cloud at least qualitatively.
The CQSM can explain this feature more qualitatively, again without
introducing additional free parameters. 
Shown in Fig.\ref{fig5} are the predictions
of the CQSM for the $\bar{d} (x) - \bar{u} (x)$ distribution as well as
the ratio $\bar{d} (x) \,/\, \bar{u} (x)$ in comparison with the corresponding
experimental data from the Hermes, FNAL-E866/NuSea group
as well as the old data from NA51.
We can say that the CQSM reproduce the characteristic features
of the empirical observations fairly well at least qualitatively.  

\unskip

\begin{figure}[h]
\begin{centering}
\includegraphics[width=11.0cm]{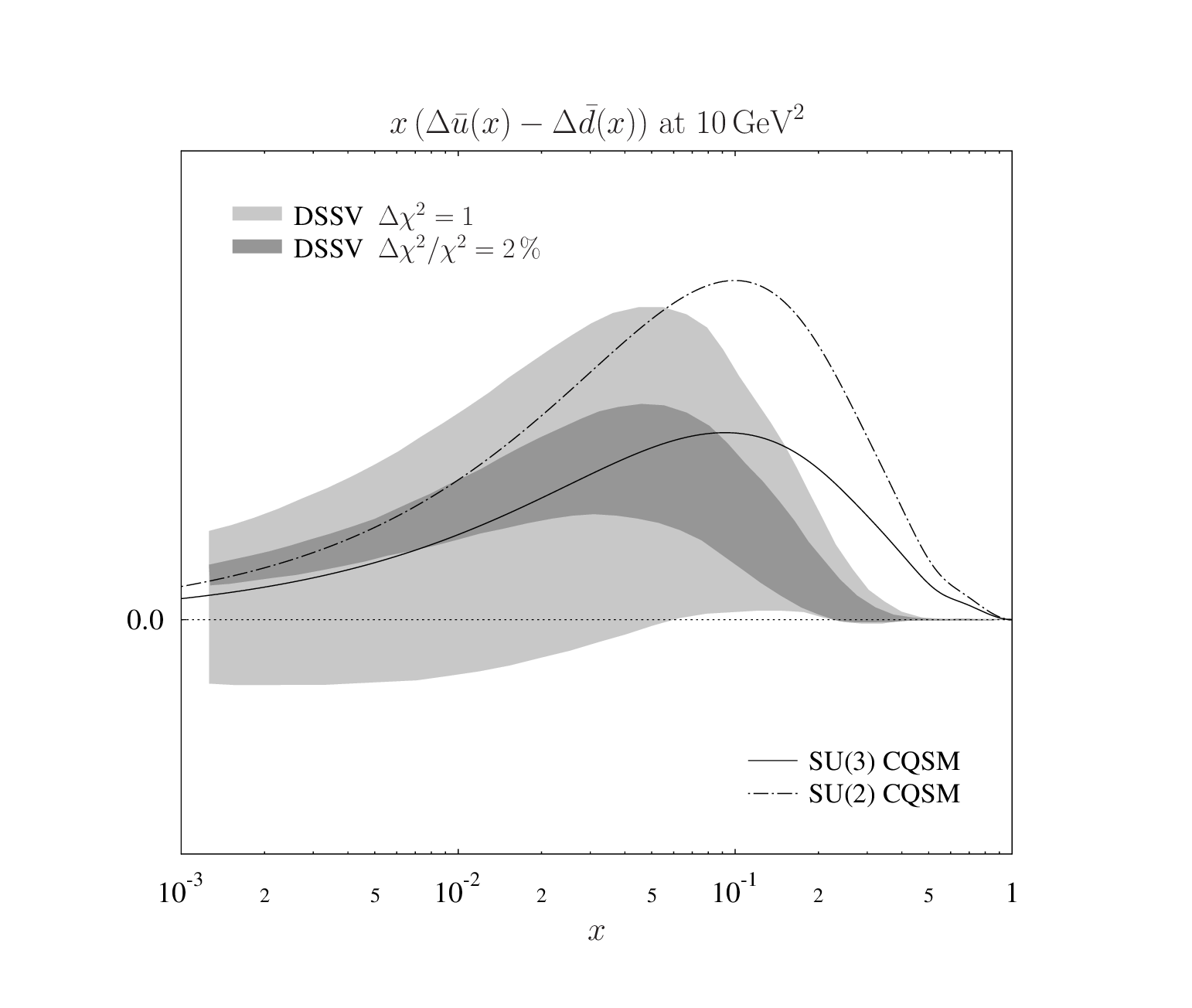}
\caption{The CQSM prediction for the flavor asymmetry of the longitudinally 
polarized anti-quark distribution $x \,( \Delta \bar{u} (x) - \Delta \bar{d} (x))$
in comparison with the empirical DSSV fit \cite{DSSV2009}.
For reference the prediction of the flavor SU(3) version of the CQSM is
also shown. \label{fig6}}
\end{centering}
\end{figure}

\vspace{4mm}
Very interestingly, the CQSM predicts the flavor asymmetry also
for the longitudinally polarized sea-quark (anti-quark) distributions.
It turns out that the model predicts that $\Delta \bar{u} (x)$ dominates over
$\Delta \bar{d} (x)$. The flavor asymmetry of the longitudinally
polarized anti-quark distributions are not yet firmly established with
the same accuracy as the flavor asymmetry of the unpolarized anti-quark
distributions. Here, in Fig.\ref{fig6}, we make a preliminary comparison
of the predictions of the CQSM with the empirical DSSV fit \cite{DSSV2009}.
One sees that the qualitative agreement between the theory and the 
empirical fit is encouraging.
Although we cannot show more examples because of the limitation of space, 
we can say with confidence that the QCSM provides us with a reliable basis 
to investigate the internal substructure of the nucleon.

\unskip

\vspace{2mm}

\section{Nucleon scalar charge density predicted by the chiral quark soliton model}

\begin{figure}[h]
\begin{centering}
\includegraphics[width=9.0 cm]{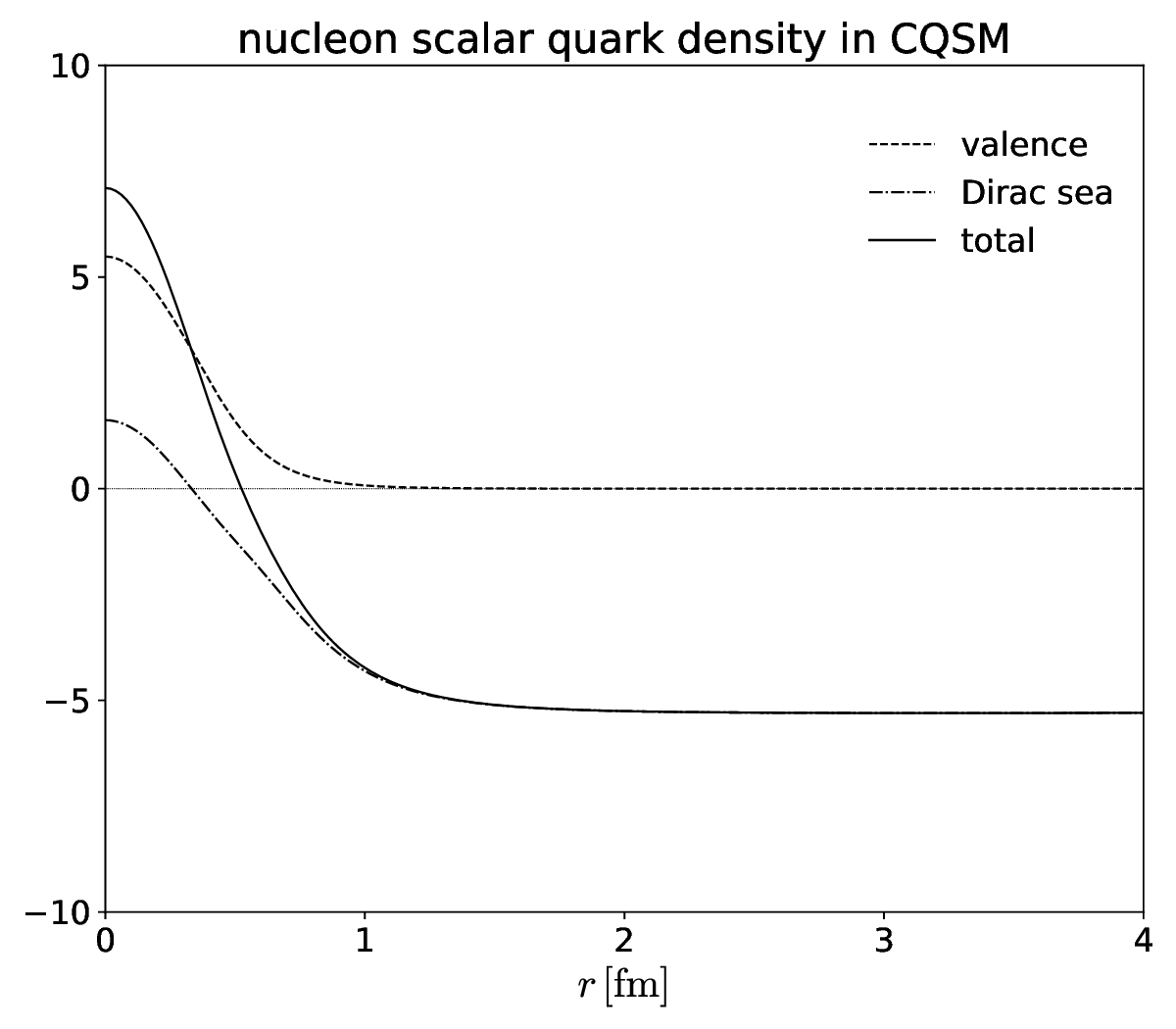}
\caption{Prediction of the CQSM for the nucleon scalar charge
density in the coordinate system. The dashed and dash-dotted curves respectively
stand for the contribution of the three quarks in the valence level and that
of the negative energy Dirac-sea quarks. \label{fig7}}
\end{centering}
\end{figure}

\vspace{1mm}
Shown in Fig.\ref{fig7} is the prediction of the CQSM for the nucleon scalar density
in the coordinate space \cite{KWW1999, RW2012}.  
(Note that the nucleon scalar charge density in Fig.\ref{fig7} is normalized as
$4 \pi \,\int_0^\infty \,[ \rho_S (r) - \rho_S (r = \infty)] \,r^2 \,d r = \bar{\sigma}$,
with $\bar{\sigma}$ being the nucleon scalar charge. )
As one sees, the contribution of the three valence
quarks smoothly attenuates to zero as the distance from the nucleon center
becomes large as is the case with the prediction of the naive three quark model.
Remarkably, however, the contribution of the negative energy Dirac-sea
quarks does not attenuate to zero but it approaches a negative non-zero value,
which is nothing but the value of the vacuum quark condensate in the QCD
vacuum. (We recall that the effective action of the CQSM was constructed
so as to reproduce the vacuum quark condensate of the QCD vacuum.)
This confirms that the CQSM can explain the vacuum quark condensate
and the nontrivial local structure of the nucleon scalar charge density at the
same time  \cite{KWW1999, RW2012}.  A question is whether this highly nontrivial 
behavior of the nucleon scalar density, i.e.
\begin{equation}
 \bar{\sigma} (r) \ \equiv \ \langle N \,\vert \bar{\psi} (\bm{r}) \,\psi (\bm{r}) \rangle_r
 \ \stackrel{r \rightarrow \infty}{\rightarrow} \ \mbox{nonzero constant} ,
\end{equation}
appear in some observables ?

\unskip

\vspace{4mm}
Note that the Fourier transform of a constant gives a Dirac's delta function.
This implies that nonzero vacuum condensate contained in the nucleon scalar 
charge density in coordinate space may appear as a delta-function singularity
in the scalar charge density (form factor) in momentum space.
Unfortunately, the Fourier transform of the local scalar charge density of the 
nucleon would not correspond to any direct observables.
As we shall see below, the relevant quantity here is the nucleon scalar charge density 
as a function of the a momentum variable $x$ of Bjorken or Feynman. 
It is the chiral-odd twist-3 quark distribution function customarily denoted as
$e (x)$

\vspace{2mm}
\section{Twist-3 PDF $e (x)$ as a nucleon scalar density in a momentum space}

\vspace{1mm}
As shown in Table 1, up to the twist-4 order, there are nine independent
quark distribution functions (See \cite{KT1999}, for example.) 
Hereafter, we call them parton distribution functions or simply PDFs.

\begin{table}[h]
\begin{center}
\begin{tabular}{|c|c|c|} \hline
 twist-2 \ & \ twist-3 \ & \ twist-4 \\ \hline\hline
 \ \ \ $f_1 (x) = q (x)$ \ \ \ & \ \ \ $e (x)$ \ \ \ & \ \ \ $f_4 (x)$ \ \ \ \\ \hline
 \ \ \ $g_1 (x) = \Delta q (x)$ \ \ \ & \ \ \ $h_2 (x)$ \ \ \ & \ \ \ $g_3 (x)$ \ \ \ \\ \hline
 \ \ \ $h_1 (x) = \Delta_T q (x)$ \ \ \ & \ \ \ $g_T (x)$ \ \ \ & \ \ \ $h_3 (x)$ \ \ \ \\ \hline
\end{tabular}
\end{center}
\caption{Nine independent quark distribution functions with twist 2, 3, and 4} 
\end{table}

\noindent
For example, $f_1 (x)$ or $q (x)$ is the familiar unpolarized PDF of the nucleon,
$g_1 (x)$ or $\Delta q (x)$ is the longitudinally-polarized PDF, and
$h_1 (x)$ or $\Delta_T q (x)$ is the so-called transversity distribution of the nucleon.
Of our particular interest here is $e (x)$, which is cllasified into a chiral-odd twist-3 PDF.
Why is it interesting ? The reason is twofold. First, the 1st moment of $e (x)$
(i.e., its integral over the Bjorken or Feynman variable $x$) gives the nucleon scalar
charge, which is proportional to the pion-nucleon sigma term.
Second, possible existence of Dirac's delta-function type singularity in $e (x)$ was
already suggested by Koike and Burkardt within the framework of perturbative
QCD \cite{BK2002}. (See also \cite{ES2003}.)
Unfortunately, the physical origin of this delta-function singularity is
not fully understood within the framework of perturbative QCD. However,
as we have already suggested, highly non-trivial structure of the nucleon
scalar density predicted by the CQSM might generate a delta-function
singularity in the scalar charge density in some momentum space.
The correctness of this expectation was shown independently in the paper
by Schweitzer \cite{Schweitzer2003} and that by ourselves \cite{WO2003}.
In these papers, it was shown that the nonperturbative origin of the
delta-function singularity in $e (x)$ can be traced back to infinite-range quark-quark
correlation of scalar type in the nucleon and that the existence of this infinite-range
correlation is inseparably connected with the nontrivial  vacuum structure
of QCD, i.e. the spontaneous $\chi$SB and the resulting non-zero vacuum
quark condensate. One might wonder why the vacuum property comes
into a hadron observable.  
As already pointed out, it is related to the  previously mentioned 
extraordinary nature of scalar quark density of the nucleon, which lives
in the nontrivial QCD vacuum.

\vspace{2mm}
Incidentally, a recent paper \cite{Ma2020} by Ma and Zhang attracted a 
renewed interest on
the existence or non-existence of the delta-function type singularity
in the twist-3 PDF $e (x)$.  According to them, within the framework
of perturbative QCD, the delta-function type singularities certainly exist but
they cancel among themselves. 
Soon after, however, their conclusion was criticized in the papers by
Bhattacharya et al. \cite{Bhattacharya2020} and also by Hatta and Zhao \cite{Hatta2020}. 
They argued that the treatment by Ma and Zhang is not justified because it neglects the 
light-front (LF) zero mode within the framework of the LF quantization,
which is vital for describing the nonperturbative vacuum of QCD.  
In any case, it is clear that the perturbative QCD may be able to predict
the existence of the delta-function singularity in $e (x)$ but it has no ability
to predict the proportionality constant of this delta-function term.
We absolutely need some nonperturbative framework like lattice QCD or
some skillfully crafted effective theory of the nucleon which takes account of the 
non-trivial vacuum structure of our real world.

\begin{figure}[h]
\begin{centering}
\includegraphics[width=6.0 cm]{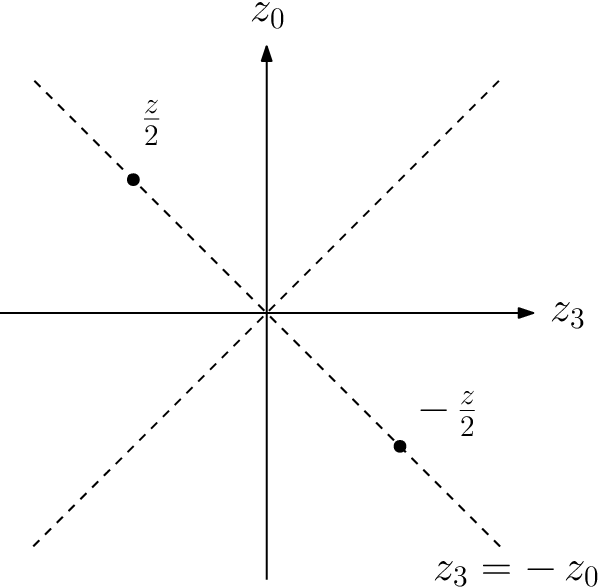}
\vspace{2mm}
\caption{Schematic figure showing the two points separated on the light-cone.\label{fig8}}
\end{centering}
\end{figure}

\vspace{2mm}
At this point, it is useful to recall the theoretical definition of the chiral-odd twist-3 
PDF $e (x)$ given as
\begin{equation}
 e (x) \ = \ M_N \,\int_{- \,\infty}^\infty \,\frac{d z_0}{2 \,\pi} \,\,
 e^{\,- \,i \,x \, M_N \,z_0} \,E (z_0) ,
\end{equation}
with
\begin{equation}
 E (z_0) \ = \ \left. \langle N \,\vert \,\bar{\psi} \left( - \,\frac{z}{2} \right) \,
 \psi \left( \frac{z}{2} \right) \,\vert N \rangle \right\vert_{z_3 = - z_0, \,z_\perp = 0} .
 \label{Eq:LC_Corr_E}
\end{equation}
That is, the PDF $e (x)$ is given as a Fourier transform of the correlation
function $E (z_0)$ of the nucleon, which measures the light-cone (LC) quark-quark
correlation of scalar type in the physical nucleon. (See Fig.\ref{fig8} for the meaning 
of the coordinate $z$.)

\unskip

\vspace{4mm}
The existence of the delta-function singularity in $e (x)$
implies the following behavior of the correlation function $E (z_0)$ : 
\begin{equation}
  E (z_0) \ \stackrel{z_0 \rightarrow \infty}{\longrightarrow} \ \mbox{non-zero constant},
\end{equation}
i.e. the existence of an infinite-range LC quark-quark correlation of scalar type.
We have already shown that, well outside the nucleon, its (local) scalar charge density 
approaches nonzero value of vacuum quark condensate.
However, what we want to really know here is the asymptotic behavior of the 
non-local quark-quark correlation with LC separation, i.e. nonlocal quark-quark 
correlation specified by Eq.(\ref{Eq:LC_Corr_E}).
Because of the limitation of the CQSM, which utilizes discretized basis for solving 
the Hartree problem \cite{KR1984, WY1991}, this $E (z_0)$ turns out to be a rapidly 
fluctuating function of  $z_0$.
It is therefore convenient to treat the corresponding smeared 
function defined as 
\begin{equation}
 \tilde{E}_\gamma (z_0) \ = \ 
 \frac{1}{\gamma \,\sqrt{\pi}} \,
 \int \,e^{\,- \,(z_0 - z)^2 \,/\,\gamma^2}
 \,E (z) \,d z ,
\end{equation}
with a suitable choice of the smearing parameter $\gamma$, which we choose
here as $\gamma = 0.05$.
For the sake of comparison, we also consider the corresponding correlator of the 
familiar unpolarized PDF,
\begin{equation}
 F(z_0)
 \ = \ \langle N \vert \,
 \bar{\psi} \left(- \frac{z}{2} \right)
 \, \gamma^+ \,\psi \left(\frac{z}{2} \right) 
 \,\vert N \rangle \,
 \vert_{z_3 = - z_0, \, z_\perp = 0} ,
\end{equation}
or its smearing version
\begin{equation}
 \tilde{F}_\gamma (z_0) \ = \ 
 \frac{1}{\gamma \,\sqrt{\pi}} \,
 \int \,e^{\,- \,(z_0 - z)^2 \,/\,\gamma^2}
 \,F (z) \,d z .
\end{equation}

\begin{figure}[h]
\begin{centering}
\includegraphics[width=8.0 cm]{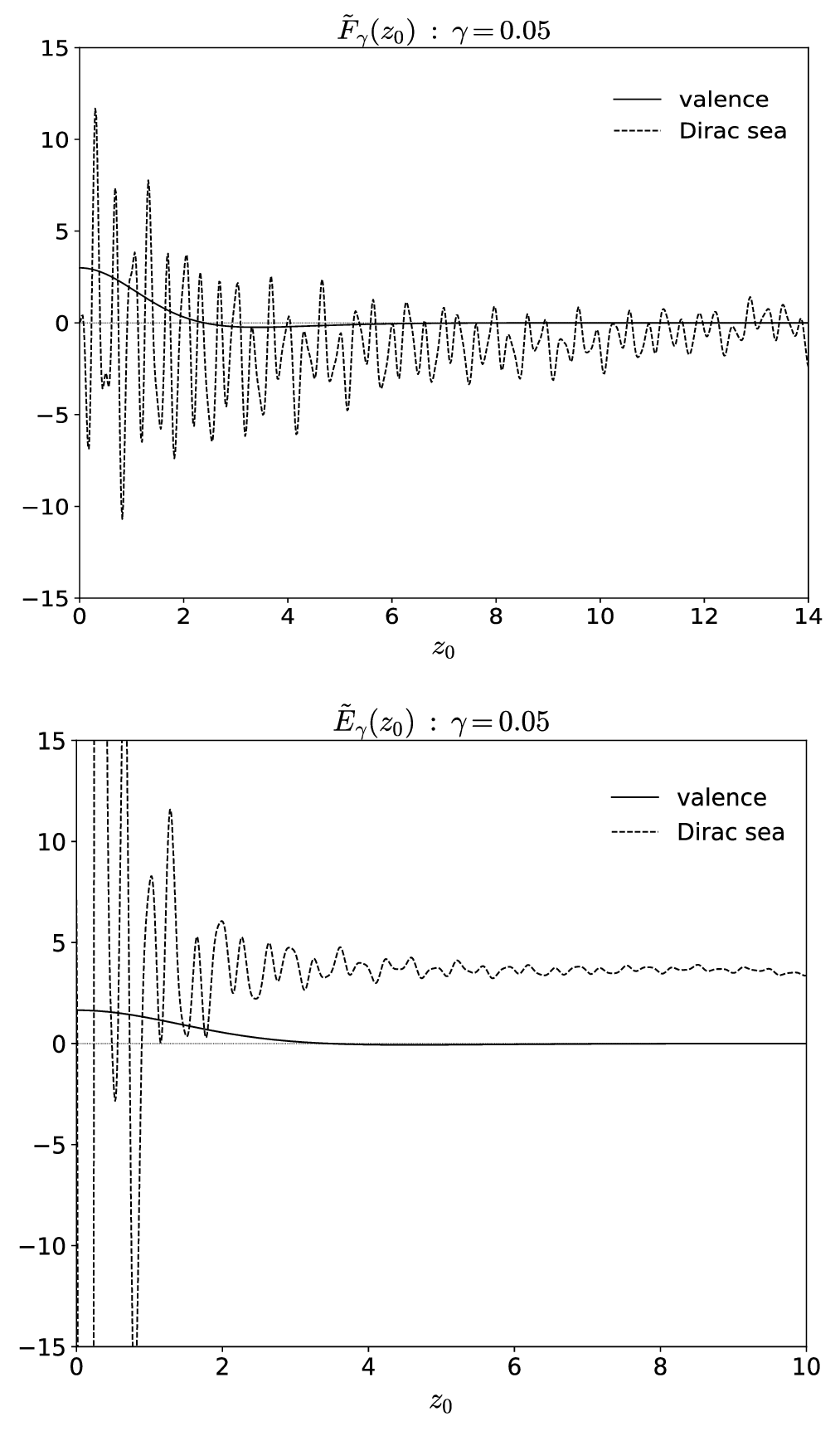}
\caption{The upper panel here shows the behavior of the smeared distribution 
$\tilde{F}_\gamma (z_0)$ corresponding to the light-cone correlator of the 
unpolarized PDF $f (x)$, while the lower panel represents the behavior of 
the light-cone correlator $\tilde{E}_\gamma (z_0)$ corresponding to the 
twist-3 PDF $e (x)$. \label{fig9}}
\end{centering}
\end{figure}

\vspace{4mm}
The upper panel in Fig.\ref{fig9} shows the smeared distribution $\tilde{F}_\gamma (z_0)$
corresponding to the correlator of the unpolarized PDF $f (x)$, while the lower panel represents
that corresponding to the smeared distribution $\tilde{E}_\gamma (z_0)$ corresponding
to the correlator of the twist-3 PDF $e (x)$.
The dashed curves in both figures show the contribution of the three 
quarks in the valence level, while the solid curves represent the contributions
of the quarks in the deformed negative energy Dirac sea orbits in the mean field.
For both distributions, the contributions of the valence quarks denoted by the solid
curves are seen to smoothly attenuate as the parameter $z_0$ as a measure of the light-cone
distance becomes larger.
In sharp contrast, there is a remarkable difference between the contributions
of the Dirac sea quarks to the correlator $\tilde{F}_\gamma (z_0)$ of the 
unpolarized PDF $f (x)$ and to the correlator $\tilde{E}_\gamma (z_0)$ of $e (x)$.
In spite of the artificial fluctuation behavior arising from the approximate
treatment by using the discretized basis, one can clearly see that the Dirac sea
contribution to $\tilde{F}_{\gamma} (z_0)$  approaches to zero as $z_0$ is
increased. On the other hand, as seen from the lower panel, the contribution
of the Dirac sea quarks to $\tilde{E}_\gamma (z_0)$
approaches some non-zero constant as $z_0$ approaches infinity.

\vspace{1mm}
To summarize, the preliminary analysis in the CQSM confirms
the highly nontrivial behavior of the two types of collation functions as follows,
\begin{eqnarray}
 &\,& F(z_0) \ \ \
 \stackrel{z_0 \rightarrow  \infty}{\longrightarrow} \ \ \ 
 \hspace{10mm} 0 , \\ 
 &\,& E(z_0) \ \ \ 
 \stackrel{z_0 \rightarrow \infty}{\longrightarrow} \ \ \ 
 \mbox{\tt nonzero constant} ,
\end{eqnarray}
which makes us convince that the PDF $e (x)$ has a delta-function singularity
at $x = 0$, whereas the PDF $f (x)$ does not.

\unskip

\vspace{3mm}
An interesting question is whether more realistic lattice QCD simulation gives a
similar prediction or not. 
Unfortunately, the lattice QCD cannot directly handle the light-cone correlators
and consequently the standard or usual PDFs. 
However, instead of usual PDFs, one may consider the {\it quasi PDFs}, which are 
given as Fourier transforms of the space-like correlators. 
As follows is very brief reminder of the concept of quasi PDF, which
was first introduced in the paper by Ji \cite{Ji2013}.  (See also \cite{JZZ2013}.) 
Some important properties of the quasi PDFs are as follows :

\vspace{1mm}
\begin{itemize}
\item They are defined as Fourier transforms of nucleon matrix element of 
{\it Lorentz-frame dependent equal-time correlators} in the large nucleon momentum
limit.

\vspace{2mm}
\item They are believed to have the same infrared behaviors as the usual PDFs.

\vspace{2mm}
\item They are not Lorentz-boost invariant but can be related to the usual PDFs
through the matching procedure in the large momentum limit.

\vspace{2mm}
\item Most importantly, the quasi PDFs are tractable within the framework of the
lattice QCD, since they are related to the space-like correlators instead of 
the light-cone correlators.
\end{itemize}
To be more explicit, we already mentioned that the twist-3 PDF $e (x)$ is
obtained as a Fourier transformation of the light-cone correlator $E (z_0)$ as
\begin{equation}
 e (x) \ = \ M_N \,\int_{- \,\infty}^\infty \,\frac{d z_0}{2 \,\pi} \,\,
 e^{\,- \,i \,x \, M_N \,z_0} \,E (z_0) ,
\end{equation}
with
\begin{equation}
 E (z_0) \ = \ \left. \langle N \,\vert \,\bar{\psi} \left( - \,\frac{z}{2} \right) \,
 \psi \left( \frac{z}{2} \right) \,\vert N \rangle \right\vert_{z_3 = - z_0, \,z_\perp = 0} .
\end{equation}
The corresponding quasi PDF $e_{qs}(x)$ is defined as a Fourier transform of the
space-like correlator $E_{qs} (z_3)$ as
\begin{equation}
 e_{qs} (x) \ = \ M_N \,\int_{- \,\infty}^\infty \,\frac{d z_3}{2 \,\pi} \,\,
 e^{\,- \,i \,x \, M_N \,z_3} \,E (z_3) ,
\end{equation}
with
\begin{equation}
 E_{qs} (z_3) \ = \ \left. \langle N \,\vert \,\bar{\psi} \left( - \,\frac{z_3}{2} \right) \,
 \psi \left( \frac{z_3}{2} \right) \,\vert N \rangle \right\vert_{z_0 = 0, \,z_\perp = 0} .
\end{equation}
Since the infrared behaviors of the usual PDF and the quasi PDF are thought to be
the same, we would expect the following behavior for $E_{qs} (z_3)$ :
\begin{equation}
 E_{qs} (z_3) \ \ \ \stackrel{z_3 \rightarrow \infty}{\longrightarrow} \ \ \ 
 \mbox{non-zero constant} . 
\end{equation}
An interesting challenge is whether the lattice QCD is able to evaluate this correlator
and whether it confirms the above conjecture or not.

\section{Direct calculation of $e (x)$ within the chiral quark soliton model}

Within the framework of the CQSM, we can evaluate the PDF  $e (x)$ itself \cite{OW2004}.
In fact, it is a little easier to directly evaluate $e (x)$ than to first evaluate
the corresponding LC correlator $E (z_0)$.
Naturally, the delta-function-type singularity cannot be handled numerically, 
so that it is convenient to first consider the smeared distribution $e_\gamma (x)$ 
corresponding to $e (x)$  : 
\begin{equation}
 e_\gamma (x) \ \equiv \ \frac{1}{\gamma \sqrt{\pi}} \,
 \int_{- \,\infty}^{+ \,\infty} \,
 e^{- \,(x - x^\prime)^2 \,/\,\gamma^2} \,
 e (x^\prime) \, d x^\prime .
\end{equation}
Note that a delta-function piece, which we expect is contained in $e (x)$,  would appear 
as a Gaussian function in the smeared distribution $e_\gamma (x)$ with the width $\gamma$.
\begin{equation}
 e_\gamma (x) \ = \ c
 \,\,\,
 \frac{1}{\gamma \sqrt{\pi}} \,
 \, e^{- \,x^2 \,/\,\gamma^2} \,
 \ \ \ \Longleftrightarrow \ \ \  
 e (x) \ = \ c 
 \,\,\delta (x) .
\end{equation}

\begin{figure}[h]
\begin{centering}
\includegraphics[width=9.0 cm]{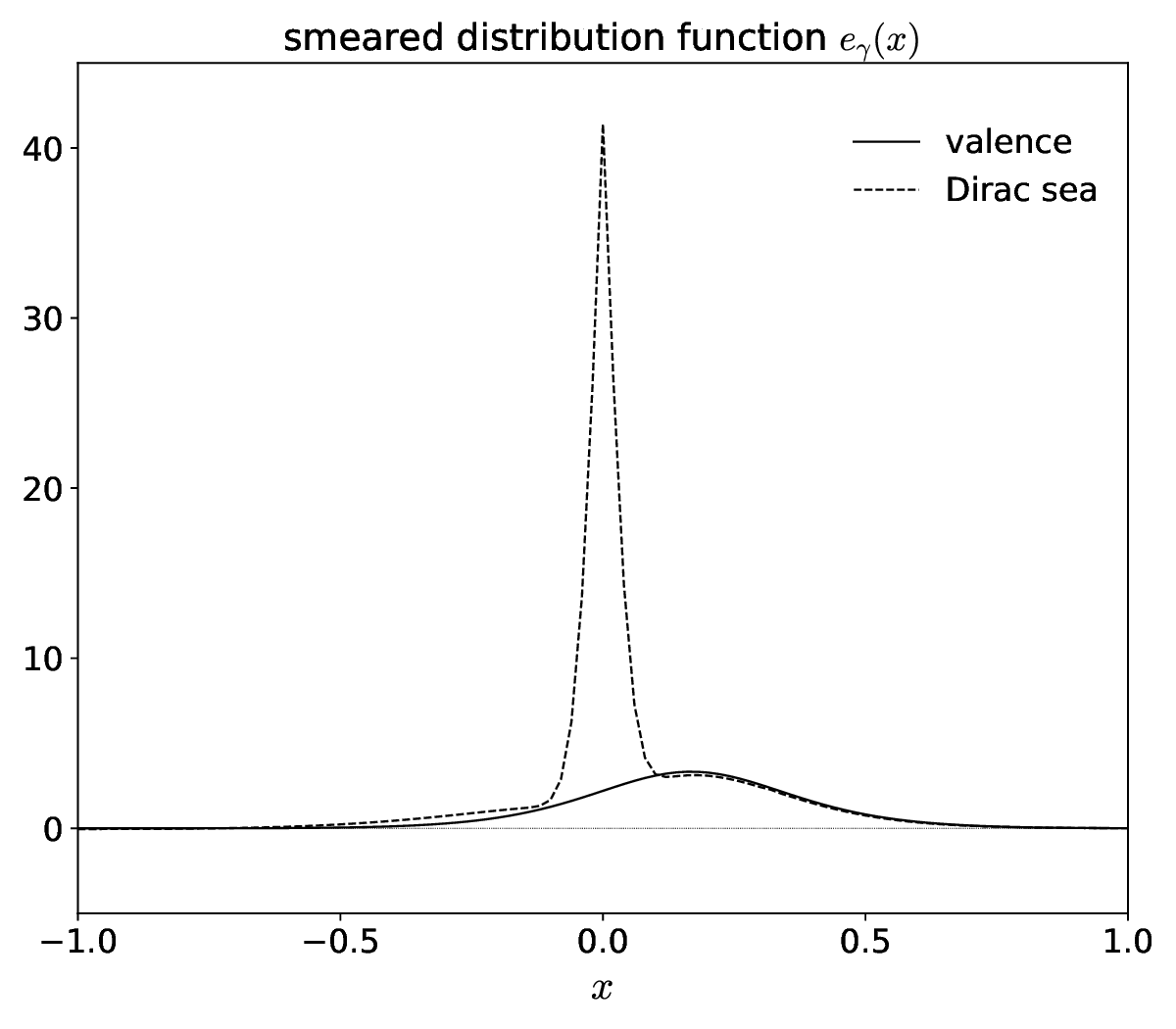}
\caption{Prediction of the CQSM for the smeared distribution function
$e_\gamma (x) \equiv e^{(T=0)}_\gamma (x)$ with a trial choice of the 
smearing parameter $\gamma = 0.06$. \label{fig10}}
\end{centering}
\end{figure}   
\unskip

\vspace{3mm}
A sample result for $e_\gamma (x) \equiv e^{(T=0)}_\gamma (x)$ corresponding to
the choice of the parameter $\gamma = 0.06$ is shown in Fig.\ref{fig6}.
(Here the superscript $(T=0)$ means the isoscalar combination for quark
flavors, i.e. $e^{(T=0)} (x) \equiv e^u (x) + e^d (x)$. We have attached this superscript, 
since, in the following, we also consider the isovector combination for quark 
flavors, i.e. the PDF $e^{(T=1)} (x) \equiv e^u (x) - e^d (x)$.)
In Fig.\ref{fig10}, the solid curve represents the contribution of  three quarks
in the valence level, whereas the dashed curve does the contribution of
the  Dirac-sea quarks. One clearly sees a peak around $x = 0$, the widths of 
which is the order of $\gamma$ coming from the Dirac-sea contribution, 
although it deviates from the expected Gaussian shape. 
\footnote{
The reason of deviation from the expected Gaussian form might need
some explanation.
We said that the vacuum polarization contribution (or the Dirac sea 
contribution) to the PDF $e(x)$ is obtained by summing over the contributions 
of all the negative-energy Dirac-sea orbitals in the hedgehog mean field.
This way of evaluating the vacuum polarization contributions is 
called the calculation based on the "occupied form". 
Alternatively, the vacuum polarization contribution can be calculated 
by summing over the contributions of all the positive-energy
Dirac-continuum (although actually discretized) orbitals. 
This way of evaluating the vacuum polarization contributions is 
called the calculation based on the "unoccupied form". 
Formally, it was proved that these two ways of calculation should 
give the same answer and it was in fact verified to be true in the
calculations of usual low energy observables. 
Unfortunately, there is some difficulty in the calculation of $x$-dependent 
PDF $e(x)$. To calculate the vacuum polarization contribution to $e(x)$ in the 
positive $x$ region, we have used the "occupied form",  while to calculate
that of $e(x)$ in the negative energy region, we have used the
"unoccupied form". The reason is because it is an effective way
to obtain $e(x)$ in each region with better numerical precision.
Unfortunately, due to the likely existence of the delta-function singularity
in $e(x)$ at $x = 0$ as well as the truncation of the discretized Kahana-Ripka
basis, this fails to precisely reproduce the expected Gaussian form in the
smeared distribution. This is not a serious problem, however, because our
demonstration here is to qualitatively convince that the $\delta (x)$-like singularity
in $e (x)$ is most likely to exist. (See the following discussion.)
}
Furthermore, we confirmed that, as the smearing parameter 
is made smaller and smaller, the width of the Gaussian-like peak gradually 
decreases and the peak eventually disappears when $\gamma$ becomes
smaller than some critical value.
This is only natural, since the delta-function cannot be reproduced exactly
with the superposition of the truncated discretized basis functions. 

\vspace{2mm}
Because of the extraordinary behavior explained above, the numerical calculation
of $e (x)$ need fairly sophisticated treatment as explained in \cite{OW2004}.
As follows are concise summary of this procedure.
We start with evaluating the smeared distribution function $e_\gamma (x)$
by using moderate value of the smearing parameter $\gamma$, which reproduces
the Gaussian-like peak around $x = 0$. Next, we continue the calculation
by gradually decreasing the magnitude of the smearing parameter $\gamma$.
We then confirm that, as $\gamma$ is decreased, the width of the Gaussian-like
peak gradually decreases, and, at the same time, the fluctuating behavior of
the Dirac-sea contribution gradually increases.  As the value of $\gamma$ is further
decreased to reach some  critical value, we see that the Gaussian-like peak disappears.
Since the remaining Dirac-sea contribution shows a fluctuating behavior with the choice
of small value of $\gamma$, we fit it by an appropriate a smooth function and then
we identify it as the regular contribution from the Dirac-seas, since its singular 
contribution corresponding to the delta function at $x = 0$ has already escaped 
from the numerical simulation. 

\begin{figure}[h]
\begin{centering}
\includegraphics[width=8.5 cm]{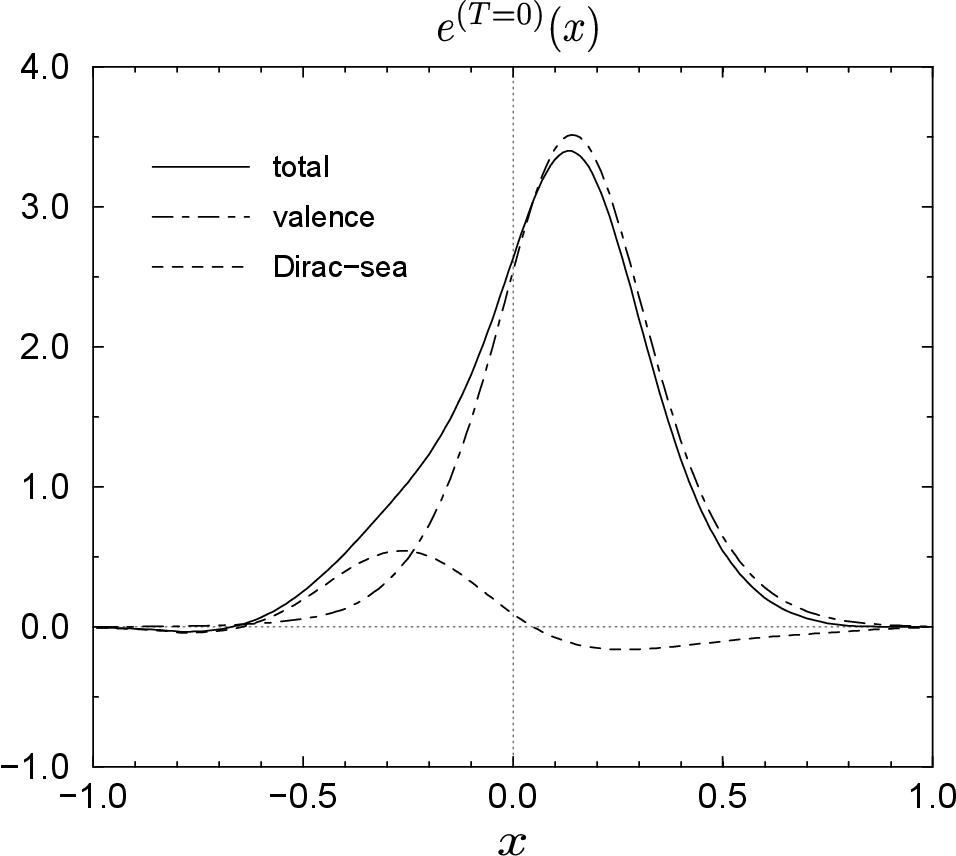}
\caption{Final prediction of the CQSM for the isoscalar combination of
the twist-3 PDF $e (x)$. The dash-dotted and dashed curves here represent
the contribution of the three valence quarks and that of the deformed
Dirac-sea quarks, while their sum is represented by the solid curves. \label{fig11}}
\end{centering}
\end{figure}   
\unskip

\vspace{4mm}
Shown in Fig.\ref{fig11} is the final numerical prediction of the CQSM for the
chiral-odd twist-3 PDF $e^{(T=0)} (x)$ with the isoscalar combination, i.e. $u + d$.
The final prediction for $e^{(T=0)} (x)$ is given as a sum of  the contribution from 
the three valence quarks and that from the Dirac-sea quarks as
\begin{equation}
 e^{(T=0)} (x) \ = \ e^{(T=0)}_{valence} (x) \ + \ e^{(T=0)}_{sea} (x) .
\end{equation}
The contribution of the Dirac-sea quarks is further divided into the singular 
contribution, which is proportional to the Dirac Delta-function $\delta (x)$ as
\begin{equation}
 e^{(T=0)}_{sea, singular} (x) \ = \ C \, \delta (x) ,
\end{equation}
and the regular contribution as 
\begin{equation}
 e^{(T=0)}_{sea} (x) \ = \ e^{(T=0)}_{sea, singular} (x) \ + \ e^{(T=0)}_{sea, regular} (x) .
\end{equation}
The regular contribution can be obtained in the numerical procedure as explained
above. However, the singular contribution cannot be obtained by the
above-explained  method, because the delta-function piece escapes from the above numerical
procedure. The question is therefore how to extract the proportionality constant $C$
above of the delta-function term.  Here, we make use of the fact that the 1st moment of 
$e^{(T=0)} (x)$ gives the nucleon scalar charge $\bar{\sigma}$.
Different from the PDF $e^{(T=0)} (x)$, the scalar charge $\bar{\sigma}$ can be evaluated
very precisely within the framework of the CQSM.
As a general rule in the CQSM, the nucleon scalar charge is also given as the
sum of the valence quark contribution and the Dirac-sea contribution as
\begin{equation}
 \bar{\sigma} \ = \ \bar{\sigma}_{valence} \ + \ \bar{\sigma}_{sea} 
 \ \simeq \ 1.7 \ + \ 10.0 \ \simeq \ 11.8. 
\end{equation}
Here, we recall the fact that the Dirac-sea contribution to $\bar{\sigma}$
is related to the 1st moment of the Dirac-sea contribution $e^{(T=0)}_{sea} (x)$ as
\begin{equation}
 \bar{\sigma}_{sea} \ = \ \int_{- \,1}^1 \,\left\{ e^{(T=0)}_{sea, singular} (x) \ + \ 
 e^{(T=0)}_{sea, regular} (x) \right\} \,d x .
\end{equation}
Since the regular part of $e^{(T=0)} (x)$, i.e. $e^{(T=0)}_{sea, regular} (x)$ was already
obtained by the numerical procedure explained above, its integral can be numerically
evaluated without any problem and the answer is given by
\begin{equation}
 \int_{- \,1}^1 \,e^{(T=0)}_{sea, regular} (x) \,d x \ \simeq \ 0.18 .
 \end{equation}
Next, after subtracting this regular contribution from the net Dirac-sea contribution
to $\bar{\sigma}$, we find that
\begin{equation}
  \int_{- \,1}^1 \,e^{(T=0)}_{sea, singular} (x) \,d x \ \simeq \ 9.92 ,
\end{equation}
which allows us to determine the proportionality constant $C$ as
\begin{equation}
 C \ \simeq \ 9.92 .
\end{equation}

\vspace{0mm}
Incidentally, with the use of the reasonable value of the average up and down quark mass
given by $m_q \simeq (4 \sim 7) \,\mbox{MeV}$, the prediction of the CQSM for
the pion-nucleon sigma term is given by
\begin{equation}
 \Sigma_{\pi N} \ = \ m_q \,\bar{\sigma} \ \simeq \ (47 \sim 83 ) \,\mbox{MeV},
\end{equation}
which seems to favor fairly large value of the pion-nucleon sigma term that is
consistent with the low-energy phenomenology \cite{Bernard1996}.
In any case, it is interesting to see that, in the CQSM, the dominant contribution 
to nucleon scalar charge comes from the contribution of the Dirac-sea quarks, 
especially from the singular delta-function term in the corresponding
PDF $e (x)$. 

\begin{figure}[h]
\begin{centering}
\includegraphics[width=9.0 cm]{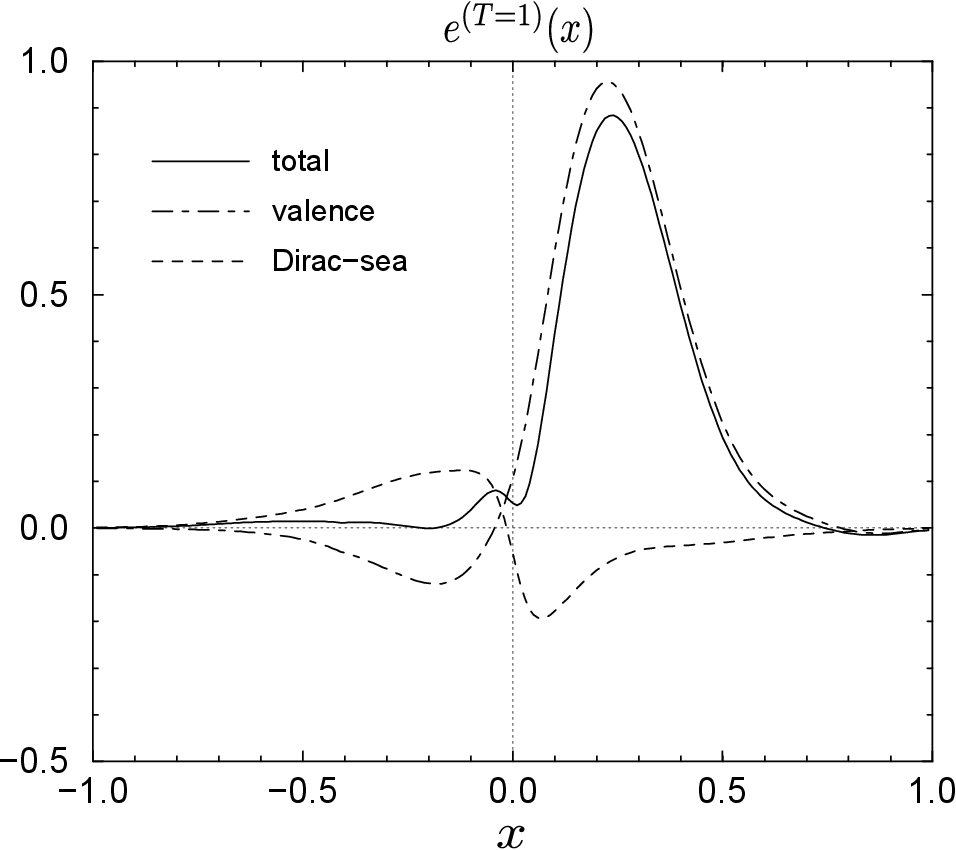}
\caption{Final prediction of the CQSM for the isovector combination of
the twist-3 PDF $e (x)$. The dash-dotted and dashed curves here respectively stand for
the contribution of the three valence quarks and that of the deformed
Dirac-sea quarks, while their sum is represented by the solid curves. \label{fig12}}
\end{centering}
\end{figure}

\vspace{2mm}
We can also calculate the isovector combination of $e (x)$, i.e.
$e^{(T=1)} (x) \equiv e^u (x) - e^d (x)$. Its calculation is much easier than that
of the isoscalar piece, because it does not contain delta-function-like singular
piece, as is consistent with that fact that, in the QCD vacuum, there is no quark 
condensate with the isovector combination.
We show in Fig.\ref{fig12} the prediction of the CQSM for the isovector combination of
the twist-3 PDF $e (x)$. Here, the contribution of the valence quarks and that of
the Dirac-sea quarks are respectively shown by the dash-dotted and dashed
curves, while the total contribution is shown by the solid curves.

\unskip

\vspace{2mm}
Combining the isoscalar- and isovector-parts of $e (x)$, we can make a
flavor decomposition and get any of
\begin{equation}
 e^u (x), \ e^d (x), \ e^{\bar{u}} (x), \ e^{\bar{d}} (x) \hspace{6mm} (x > 0)
\end{equation}
Here we have made use of the fact that, for the PDF $e (x)$,
the calculated distribution function of the quark in the negative $x$ region can be
interpreted as the distribution of the corresponding anti-quarks  according to
the rule $e^q (-x) = e^{\bar{q}} (x)$ with $0 < x < 1$.

\vspace{2mm}
\begin{figure}[h]
\begin{centering}
\includegraphics[width=8.0 cm]{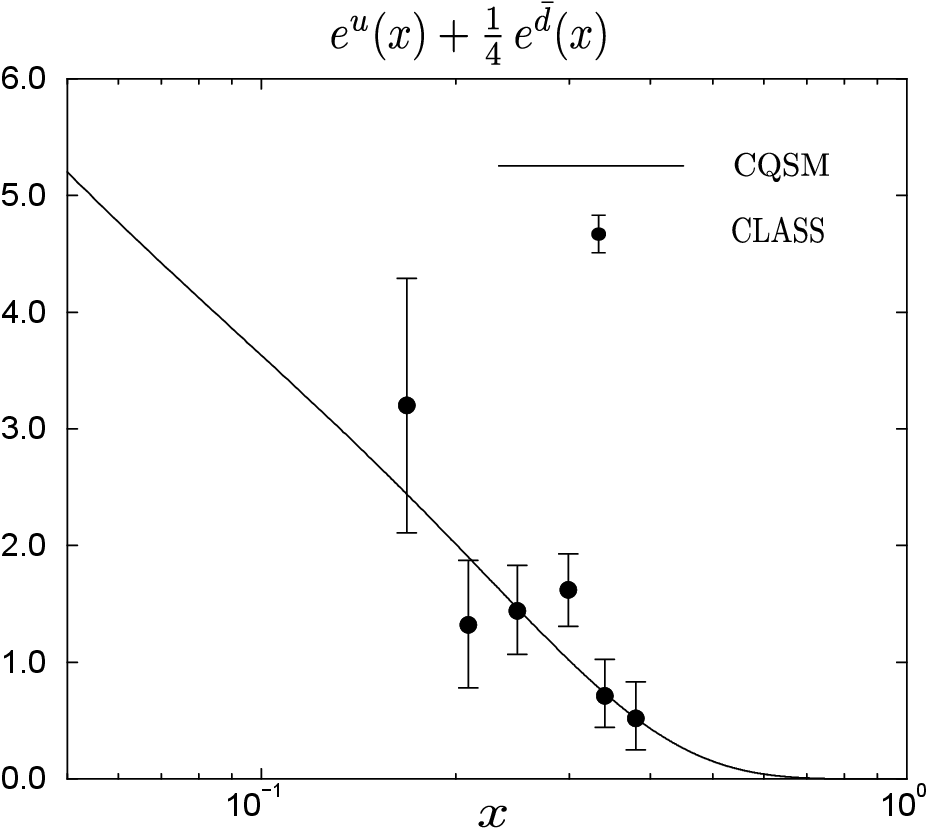}
\caption{Preliminary comparison of the CQSM for the twist-3 PDF $e (x)$
of the flavor combination $e^u (x) + \frac{1}{4} \,e^{\bar{d}} (x)$
corresponding to the energy scale $Q^2 \simeq 5 \,\mbox{GeV}$ with
the empirical data extracted from the CLAS measurement by Efremov,
Goeke and Schweitzer \cite{EGS2002}. \label{fig13}}
\end{centering}
\end{figure}   
\unskip

\vspace{5mm}
Shown in Fig.\ref{fig13} is the preliminary comparison of the prediction of the CQSM
for the twist-3 PDF $e (x)$ with the flavor combination
$e^u (x) + \frac{1}{4} \,e^{\bar{d}} (x)$ with the empirical extraction of the 
corresponding PDF at $Q^2 \simeq 5 \,\mbox{GeV}$ from the CLAS 
semi-inclusive scattering data.
Although the comparison is very preliminary, the agreement between
the theoretical prediction and the empirical data is encouraging.
We emphasize the fact that the theoretical curve here contains
only the sum of the three valence quarks contribution and the regular
part of the Dirac-sea contribution, and these contributions are
far much smaller than the singular part of the Dirac-sea contribution
which is concentrated at $x = 0$ as a Dirac delta-function.
This observation together with rough agreement between the theory
and the empirical data already appear to support the likely existence of
the delta-function singularity in $e (x)$ with sizable strength. 
Naturally, to get more confirmative evidence, more precise extraction of
the PDF $e (x)$ from the analysis of the relevant semi-inclusive processes
is mandatory especially down to the small $x$ region 
as much as possible.  
(For more recent experimental status, see \cite{Courtoy_S2022, Courtoy_Class2022}, 
for example.)
If this becomes in fact possible, we may be able to confirm the existence of
the delta-function in $e (x)$ as a signal of the nontrivial vacuum structure
of QCD even though somewhat indirectly.

\vspace{2mm}
\section{Summary}

\vspace{2mm}
The CQSM predicts fairly unusual behavior of the nucleon scalar charge densities
as follows :

\vspace{1mm}
\begin{itemize}
\item $\langle N \,|\,\bar{\psi} \,\psi
 \,|\, N \rangle_r \ \ 
 \stackrel{r \rightarrow \infty}{\longrightarrow}
 \ \ \mbox{\tt nonezero constant}$
 \vspace{2mm}
 \item existence of $\delta (x)$-type singularity in the chiral-odd twist-3 PDF $e (x)$.
\end{itemize}
These predictions of the chiral quark soliton model (CQSM) for the chiral-odd twist-3
PDF $e  (x)$ comes from its unique feature 
such that it can simultaneously describes the nontrivial vacuum quark condensate and 
the local structure of the nucleon scalar charge distribution. 
An interesting question is whether the lattice QCD simulation would confirm these 
unique predictions of the CQSM in the scalar channel,
which has the same quantum number as the physical vacuum.
As is well known,  although the light-cone PDFs cannot be handled by the 
lattice QCD framework, the corresponding quasi PDFs would in principle be tractable.
Due to the nontrivial behavior of the QCD vacuum characterized by non-zero
quark condensate, such a simulation in the scalar channel would not be
very easy, but it must be a great challenge to the framework of lattice QCD.

\vspace{2mm}
From the experimental side, the precise extraction of the PDF
$e (x)$ is not an easy task. This is because, its chiral-odd nature forbids
its measurement through the well-understood inclusive deep-inelastic scatterings (DIS).
To extract it, one must use more complicated semi-inclusive DIS processes.
Moreover, the delta-function singularity at $x = 0$ cannot be a directly
accessed through the framework of the high-energy deep-inelastic scattering
measurement. The best one can do is to extract $e (x)$ down to a
smallest possible value of $x =x_{min}$, and evaluate its integral over $x$ between 
$x_{min}$ and $1$. If the coefficient of the delta-function term is so large as predicted 
by the CQSM, this integral would significantly underestimate the value of
nucleon scalar charge as expected from the pion-nucleon sigma term sum rule.
It appears to us that several preliminary analyses already indicates the validity of this
anticipation.  

%




\acknowledgments{The present paper is based on the author's talk at Journal Club at KEK in 2021. 
The author would like to thank the members of KEK Theory Center for critical but 
constructive discussions and advises.}







\end{document}